\documentclass[11pt,twoside, letterpaper]{article}
\usepackage{amssymb}

\newcommand{\be}{\begin{equation}}
\newcommand{\ee}{\end{equation}}
\newcommand{\ba}{\begin{array}}
\newcommand{\ea}{\end{array}}
\newcommand{\p}{\partial}
\newcommand{\bu}{\mathbf{u}}
\newcommand{\F}{\mathbf{F}}

\def\diag{\mathop{\rm diag}}

\newcommand{\ds}{\displaystyle}

\newtheorem{prop}{Proposition}
\newtheorem{lem}[prop]{Lemma}
\newtheorem{cor}[prop]{Corollary}

\def\endproof{\ifmmode \qed \\\hbox{\ } 
\else $\qed$ \vskip\proclaimskip \fi}
\date{}

\title{\vspace*{-15mm}\bf On a class of inhomogeneous extensions\\ for
integrable
   evolution systems\footnote{This paper is published 
   in Proceedings of $8^{th}$ Intl.\ Conf.\ 
on Diff. Geometry and its Applications: {\em Differential
Geometry and its Applications} (Proc.\ $8^{th}$ Int.\ Conf.), Silesian
University in Opava, Opava, Czech Republic, 2001, p.243--252.}
\vspace{-3mm}}
\author{{\sc Artur Sergyeyev}\\
Silesian University in Opava,
Mathematical Institute,\\
Na Rybn\'\i{}\v cku 1, 746~01 Opava, 
Czech Republic\\
E-mail: {\tt Artur.Sergyeyev@math.slu.cz}}

\begin{document}

\maketitle
\vspace{-7mm}
\begin{abstract}
In the present paper we prove the integrability (in the
sense of existence of formal  symmetry of infinite rank) for a
class of block-triangular inhomogeneous extensions of
(1+1)-dimensional integrable evolution systems. An important
consequence of this result is the existence of formal  symmetry of
infinite rank for ``almost integrable'' systems,  recently
discovered by Sanders and van der Kamp.

\vspace{1mm}


\vspace{1mm}

\end{abstract}

%

\section*{Introduction}
It is well known that the existence of infinite number of
generalized (or higher order) symmetries for a system of PDEs is
one of the most  important signs of its integrability, see for
example \cite{fokas,s,mik1,mik,o}. Moreover, for a long time it was
generally believed that the existence of only one  nontrivial local
generalized symmetry implies the existence of infinitely many  such
symmetries, cf.\ \cite{fokas}.

However, the latter statement is not true, as shows 
the example of Bakirov
system \cite{bak91}\looseness=-1
\be\label{bakir}
\begin{array}{l}
 \partial u / \partial t=\partial^4 u / \partial x^4+v^2,\\
 \partial v / \partial t=(1/5) \partial^4 v / \partial x^4.
\end{array}
\ee
This system has only one non-Lie-point $x,t$-independent local 
generalized symmetry,
as it was proved by Beukers, Sanders and Wang \cite{bsw} using the
sophisticated  methods of number theory. What is more, the situation 
remains unchanged even if we pass to $x,t$-dependent local generalized 
symmetries, see \cite{as}.

Sanders and van der Kamp \cite{sv} have generalized this result and
found a counterexample to the conjecture of Fokas \cite{fokas}
stating that if an $s$-component system of PDEs has $s$ 
non-Lie-point  local generalized symmetries, then it has infinitely
many symmetries of this kind. Namely, they have exhibited 
a two-component
evolution system possessing only two
non-Lie-point $x,t$-independent local generalized symmetries. This
system is of the form\looseness=-1
\be\label{svk}
\begin{array}{l}
 \partial u / \partial t=a u_{7}+b v_{1}v_{2}+7 v v_{3},\\
 \partial v / \partial t=v_{7},
\end{array}
\ee
where $a=-(42\alpha^{5} +280\alpha^{4} +700\alpha^{3} +798\alpha^{2}
+504\alpha +104)$, $b= 7\alpha^{5} +49\alpha^{4} +133\alpha^{3}
+175\alpha^{2} +126\alpha +56$, and $\alpha$ is a root of the equation
$\alpha^{6} +7\alpha^{5} +19\alpha^{4} +25\alpha^{3} +19\alpha^{2}
+7\alpha+1 =0$, $u_i=\partial^i u/\partial x^i$,  $v_i=\partial^i
v/\partial x^i$.

Let us note that both systems (\ref{bakir}) and (\ref{svk}) 
are exactly solvable.  Indeed, one can find the general solution of
the second equation for $v$, then plug it into the first equation and
find its general solution for $u$.

Since the systems (\ref{bakir}) and (\ref{svk}) possess only a finite
number of non-Lie-point local generalized symmetries and at the same
time are exactly solvable, it is interesting to find out
whether they pass or fail other integrability tests. One of the most
powerful and algorithmic tests of this kind is 
the existence of nondegenerate formal symmetry
of  infinite rank and nonzero degree, see
\cite{s,mik1,mik,o} and Section~1 below for details.  For the Bakirov
system the existence of formal symmetry with these properties was
proved by Bilge \cite{bilge}. It is natural to ask whether  a similar
result could be established for the system (\ref{svk}), as well as
for other systems listed in \cite{sv}.
\looseness=-1

In the present paper we show that this is indeed possible for quite a
large class of evolution systems of the form (\ref{eveq3}), which
naturally generalize the Bakirov system~\cite{bak91},  and those of
Sanders and van der Kamp~\cite{sv}, see Proposition~\ref{prop1} and
Corollaries~\ref{cor3}--\ref{cor5} below for details. Note that, in
the terminology of Kupershmidt \cite{bk}, the system
(\ref{eveq3}) can be thought of as a particular case of {\em
inhomogeneous nonlinear extension}  of its last block, that is,
$\p \vec u^s/\p t=\vec f^s(x,t,\vec u^s,\vec u^s_{1},\dots,
\vec u^s_{n})$.

\section{Basic definitions and structures}

Consider a $(1+1)$-dimensional evolution system 
\begin{equation} \label{eveq1}
\partial \bu / \partial t =\F(x,t,\bu,\dots,\bu_{n})
\end{equation}
for the $q$-component vector function $\bu=(u^{1},\dots,u^{q})^{T}$.
Here $\bu_{j}=\p^{j}\bu/\p x^{j}$, $\bu_{0}\equiv \bu$,
$\F=(F^{1},\dots, F^{q})^{T}$, and the superscript ${T}$ denotes the
matrix transposition. In what follows we assume that $n \geq 2$ and
$\p\F/\p\bu_{n}\neq 0$.

Let us recall that a function $f$ of $x, t, \bu, \bu_{1},\dots,$ is
called {\em local} (cf.~\hbox{\cite{mik}}) if it depends only on a
finite number of variables $\bu_{j}$.
The operators of total derivatives with respect to $x$ and $t$
on the space of (smooth) local functions take the form
$D_{x}\equiv D = \p /\p x + \sum_{i=0}^{\infty} \bu_{i+1} \p/\p\bu_{i}$ 
and 
$D_{t}=\p/\p t+ \sum_{i=0}^{\infty}D^{i}(\F)\p/\p\bu_{i}$, 
cf. \cite{mik1,mik,o}.\looseness=-1

Consider \cite{mik1,mik,o} a formal series in powers of $D$ of the
form 
$$
\mathfrak{H} = \sum_{j=-\infty}^{q} h_{j}D^{j},
$$
where $h_{j}$ are $(p \times p)$-matrix-valued local functions.
The greatest $m\in\mathbb{Z}$ such that $h_{m}\neq 0$ is called
the {\em degree} of $\mathfrak{H}$ and is denoted by
$m =\deg\mathfrak{H}$. The formal series  $\mathfrak{H}$ is called
{\em nondegenerate}, if $\det h_{m}\neq 0$, $m=\deg\mathfrak{H}$. 
Following the usual convention \cite{o}, we assume that $\deg
0=-\infty$.

A formal series $\mathfrak{R}=\sum_{j=-\infty}^{r}\eta_{j}D^{j}$,
where $\eta_{j}$ are $(q\times q)$-matrix-valued local functions, is
called a {\em formal symmetry} of infinite rank
(see \cite{s,mik1,mik,o}) for (\ref{eveq1}), if it satisfies the
equation
\be \label{wfs-inf}
D_{t}(\mathfrak{R})=[\F_{*},\mathfrak{R}].
\ee
Here we set $D_{t}(\mathfrak{R})=\sum_{j=-\infty}^{r}D_{t}
(\eta_{j})D^{j}$, $\F_{*}=\sum_{i=0}^{n}\p\F/\p\bu_{i}D^{i}$,
and $[\cdot,\cdot]$ stands for the usual commutator of two formal
series:
$[\mathfrak{A}, \mathfrak{B}]=\mathfrak{A}\circ \mathfrak{B} 
- \mathfrak{B}\circ\mathfrak{A}$.

The multiplication law $\circ$ (see
for example \cite{o}) is defined for monomials as 
$$
a D^{i}\circ b D^{j} =a \sum\limits_{q=0}^{\infty}
{\ds \frac{i(i-1)\cdots
(i-q+1)}{q!}}D^{q}(b)D^{i+j-q},\quad i,j\in\mathbb{Z},
$$
and is extended by linearity to the set of all formal series. In what 
follows we shall omit $\circ$ unless this leads to confusion. In
particular,  for $k\in\mathbb{N}$ we set
$\mathfrak{R}^{k}=\mathfrak{R}\circ\mathfrak{R}^{k-1}$.




\section{The main result}
Consider an evolution system (\ref{eveq1}) of the form
\be\label{eveq3}
\ba{l}
{\ds {\frac{\p \vec u^1}{\p t}}}
=
\vec f^1(x,t, \vec u^1,\vec u^1_{1},\dots,\vec u^1_{n})
  +\vec h^1(x,t,\vec u^2, \vec u^2_{1},\dots,\vec u^2_{n-1},\dots,
           \vec u^s,\vec u^s_{1},\dots,\vec u^s_{n-1}),
           \\
{\ds\frac{\p \vec u^2}{\p t}}
=
\vec f^2(x,t,\vec u^2,\vec u^2_{1},\dots,\vec u^2_{n})
  +\vec h^2(x,t,\vec u^3,\vec u^3_{1},\dots,\vec u^3_{n-1},\dots,
            \vec u^s,\vec u^s_{1},\dots,\vec u^s_{n-1}),
            \\
 \hspace*{8,5mm}\vdots\\
{\ds\frac{\p \vec u^s}{\p t}}
=\vec f^s(x,t,\vec u^s,\vec u^s_{1},\dots,\vec u^s_{n}),
\ea
\end{equation}
where $n\geq 2$, 
$\vec u^{\alpha}_{j}=\partial^{j} \vec u^{\alpha}/\partial x^{j}$,  
$\vec u^{\alpha}=(u^{\alpha,1},\dots, u^{\alpha,q_{\alpha}})^{T}$,
$\vec f^{\alpha}=(f^{\alpha,1},\dots, f^{\alpha,q_{\alpha}})^{T}$,
$\vec h^{\alpha}=(h^{\alpha,1},\dots, h^{\alpha,q_{\alpha}})^{T}$.

The system (\ref{eveq3}) is nothing but a particular case
of (inhomogeneous nonlinear) extension of $\vec u^s_t=\vec f^s$, cf.\
\cite{bk}. It turns out that under some extra conditions 
the existence of
formal symmetries  of infinite rank and nonzero degree for the
systems
$\vec u^\alpha_t=\vec f^\alpha$, $\alpha=1,\dots,s$,  
implies the same property for the system (\ref{eveq3}) with
arbitrary~$\vec h^{\alpha}$. 
\looseness=-2

In what follows we assume the ground field to be algebraically closed, 
so that any matrix can be brought into Jordan's normal form, see e.g.\
\cite{gant}. Then we have the following result.

\begin{prop}\label{prop1}
Suppose that the matrices $\p\vec f^{\alpha}/\p\vec u_n^{\alpha}$ and 
$\p\vec f^{\beta}/\p\vec u_n^{\beta}$ have no common eigenvalues
(i.e., the eigenvalues in question are distinct as functions) 
for all $\alpha\neq\beta$, $\alpha,\beta=1,\dots,s$, and at least one
of these matrices  is nonzero.
Further assume that each of the evolution systems 
$\vec u_t^\alpha=\vec{f}^\alpha$, $\alpha=1,\dots,s$, has a formal
symmetry $\mathfrak{L}_{\alpha}$ of infinite rank and nonzero
degree,  and the coefficients of $\mathfrak{L}_{\alpha}$ for
$\alpha=1,\dots,s-1$ depend on $x$ and $t$ only. 
\looseness=-1

Then the system {\rm (\ref{eveq3})} with arbitrary (smooth) functions
$\vec h^{\alpha}(x,t,\vec u^{\alpha+1}, \vec u^{\alpha+1}_{1},
\dots, \vec u^{\alpha+1}_{n-1},\dots, \allowbreak\vec u^s, \vec
u^s_{1},\dots, \vec u^s_{n-1})$,
$\alpha=1,\dots,s-1$, also possesses a formal symmetry 
of infinite rank and nonzero degree. 

Moreover, if all $\mathfrak{L}_{\alpha}$ are nondegenerate, then
{\rm (\ref{eveq3})} possesses a {\em nondegenerate} formal symmetry 
of infinite rank and nonzero degree.
\end{prop}
\vspace{-1mm}
\hspace*{\parindent} {\bf Proof. } Let us start with the following
lemma (cf.\ \cite[Proposition~2.1]{mik1}).
\vspace{-1mm}
\begin{lem}\label{lemma}
Suppose that the matrices $\p\vec f^{\alpha}/\p\vec u_n^{\alpha}$ and 
$\p\vec f^{\beta}/\p\vec u_n^{\beta}$ have no common eigenvalues
(i.e., the eigenvalues in question are distinct as functions) 
for all $\alpha\neq\beta$, $\alpha,\beta=1,\dots,s$, and at least one
of these matrices  is nonzero.

Then there exists a unique formal series 
$$
\mathfrak{T}=\mathbf{1}+\sum_{i=-\infty}^{-1}T_{i}D^{i}
$$ 
such that $T_i$ are upper block-triangular $(q\times q)$-matrix-valued
local functions with zero diagonal blocks and  we have\looseness=-1 
$$
\mathfrak{V}\equiv \mathfrak{T}\F_{*}\mathfrak{T}^{-1} 
    + D_{t}(\mathfrak{T}) {\mathfrak{T}}^{-1}
 = \diag(\mathfrak{F}_{1},\dots,\mathfrak{F}_{s}).
$$
Here $\mathbf{1}$ is a $q\times q$ unit matrix, 
$q=\sum_{\alpha=1}^{s} q_{\alpha}$,
$\mathfrak{F}_{\alpha}=
\sum_{i=0}^{n}\p\vec{f}^{\alpha}/\p\vec{u}^{\alpha}_i
D^i$, and $\F$ stands for the right-hand side of
{\rm (\ref{eveq3})}.
\looseness=-1
\end{lem}

Before we prove this lemma, let us 
apply it for the proof
of Proposition~\ref{prop1}. Equation (\ref{wfs-inf})
under the  transformation
$\mathfrak{R} \rightarrow \mathfrak{L}=
\mathfrak{T}\mathfrak{R}\mathfrak{T}^{-1}, 
\F_{*} \rightarrow \mathfrak{V}$ becomes\looseness=-1
\be\label{trans}
D_{t}(\mathfrak{L})=[\mathfrak{V}, {\mathfrak{L}}].
\ee

For the system (\ref{eveq3}),  using the assumption 
that the matrices $\p\vec f^{\alpha}/\p\vec u_n^{\alpha}$ 
have no common eigenvalues,  it is easy to check (cf. \cite{mik1})
that the coefficients of any solution
$\mathfrak{L}$ of (\ref{trans}) are block-diagonal matrices, i.e.,
$\mathfrak{L}=\diag(\mathfrak{R}_{1},\dots,\mathfrak{R}_{s})$, where 
$\mathfrak{R}_{\alpha}$ is a formal series whose coefficients are 
$q_{\alpha}\times q_{\alpha}$ matrices, and thus (\ref{trans}) is
broken into $s$ blocks:\looseness=-1
\be\label{frmsym1}
D_{t}(\mathfrak{R}_{\alpha})=\lbrack \mathfrak{F}_{\alpha},
\mathfrak{R}_{\alpha}\rbrack.
\ee
Each of equations (\ref{frmsym1}) for $\alpha=1,\dots,s$ 
has a
solution
$\mathfrak{R}_{\alpha}=\mathfrak{L}_{\alpha}$. 
Thus, equation (\ref{trans}) has a solution
$\mathfrak{L}=\diag(\mathfrak{L}_{1},\dots,
\mathfrak{L}_{s})$, 
and $\mathfrak{R}=\mathfrak{T}^{-1}\mathfrak{L}\mathfrak{T}$, with
$\mathfrak{T}$ constructed in Lemma~\ref{lemma}, is 
a formal symmetry of
nonzero degree and infinite rank for (\ref{eveq3}).
\looseness=-2

If all $\mathfrak{L}_{\alpha}$ are nondegenerate,
then we can choose $\mathfrak{L}$ to be 
$\mathfrak{L}=\diag(\mathfrak{L}_{1}^{p_1},\dots,
\mathfrak{L}_{s}^{p_s})$ and
$\mathfrak{R}=\mathfrak{T}^{-1}\mathfrak{L}\mathfrak{T}$, where
$p_{\alpha}=m/m_{\alpha}$,
$m_{\alpha}=\deg\mathfrak{L}_{\alpha}$, and $m$ is the least common
multiple of $m_{\alpha}$, $\alpha=1,\dots,s$. Thus constructed
$\mathfrak{R}$ obviously will be a nondegenerate formal symmetry of
infinite rank and nonzero degree $m$ for (\ref{eveq3}).  This remark
completes the proof of Proposition~\ref{prop1}. 
$\square$.

Note that $D_t$ in (\ref{frmsym1}) {\em does not} coincide with 
the operator 
$\p/\p t+\sum_{i=0}^{\infty}
D^i\!(\vec f^{\alpha})\p\!/\p\vec u^{\alpha}$ 
(no sum over $\alpha$). Therefore,
if the coefficients of $\mathfrak{L}_\alpha$,
$\alpha=1,\dots,s-1$, depend not only on $x$ and $t$, there
is no obvious way to construct the solutions of (\ref{frmsym1}) for
$\alpha=1,\dots,s-1$ and to extend the result of Proposition~1 to
this case. 

{\bf Proof of Lemma 2.}
By the above,  $\mathfrak{T}$ is assumed to have the form
\[
\mathfrak{T}=\left(
\ba{ccccc} {\bf{1}}_{q_1} &\mathfrak{T}_{12} 
&\dots &\mathfrak{T}_{1s}\\
0 &{\bf{1}}_{q_2} 
&\dots &\mathfrak{T}_{2s}\\
\vdots &\vdots &\ddots &\vdots\\
0 &0  &\dots  &{\bf{1}}_{q_{s}}
\ea \right),
\]
where ${\bf{1}}_{q_{\alpha}}$ stands for $q_{\alpha}\times
q_{\alpha}$ unit matrix,
$\mathfrak{T}_{\alpha\beta}$ are formal series of degree not higher
than $-1$: $\mathfrak{T}_{\alpha\beta}
=\sum_{r=1}^{\infty}\tau_{\alpha\beta}^{r}D^{-r}$, and the
coefficients $\tau_{\alpha\beta}^{r}$  are $q_{\alpha}\times
q_{\beta}$ matrices.

It is clear that for $\F$ being the right-hand side of (\ref{eveq3})
$\F_*$ has a similar structure:
$$
\F_*\equiv\mathfrak{V}+\mathfrak{B}=\left(
\ba{cccc} \mathfrak{F}_{1} &\mathfrak{B}_{12} 
&\dots &\mathfrak{B}_{1s}\\
0 &\mathfrak{F}_{2} 
&\dots &\mathfrak{B}_{2s}\\
\vdots &\vdots &\ddots &\vdots\\
0  &0    &\dots &\mathfrak{F}_{s}
\ea
\right),
$$
where $\mathfrak{B}_{\alpha\beta}$ are formal series of degree 
not higher than $n-1$:
$$
\mathfrak{B}_{\alpha\beta}=\sum
_{r=0}^{n-1}b_{\alpha\beta}^{r}D^{r},
$$ 
and the coefficients $b_{\alpha\beta}^r$ are 
$q_{\alpha}\times q_{\beta}$ 
matrices.

Multiplying the equality $\mathfrak{V}=
\mathfrak{T}\F_{*}\mathfrak{T}^{-1} + D_{t}(\mathfrak{T})
{\mathfrak{T}}^{-1}$ by $\mathfrak{T}$ on the right,
we find
$$
\mathfrak{V}\mathfrak{T}=\mathfrak{T}\F_{*}+ D_{t}(\mathfrak{T}).
$$
Inserting in this formula the expressions for $\mathfrak{T}$,
$\mathfrak{V}$ and $\F_{*}$ and equating ``blockwise''
its left-hand side and right-hand side, we obtain identities of the
form $\mathfrak{F}_{\alpha}=\mathfrak{F}_{\alpha}$ or $0=0$ together
with the following equations:
\be\label{p1}
\mathfrak{F}_{\alpha}\mathfrak{T}_{\alpha\beta}
-\mathfrak{T}_{\alpha\beta}\mathfrak{F}_{\beta}=
\sum\limits_{\gamma=\beta+1}^{s}\mathfrak{T}_{\alpha\gamma}
\mathfrak{B}_{\gamma\beta}
+\mathfrak{B}_{\alpha\beta}+D_{t}(\mathfrak{T}_{\alpha\beta}), 
\quad s\geq\beta>\alpha\geq 1.
\ee

Provided the coefficients of formal series 
$\mathfrak{T}_{\alpha\gamma}$,
$\gamma>\beta$,
are known, we can find from (\ref{p1}) the coefficients of
$\mathfrak{T}_{\alpha\beta}$, solving {\em algebraic} equations only.
Indeed, equating the coefficients at $D^{n-p}$ on the
left- and right-hand side of (\ref{p1}) yields 
$$
 a^{\alpha}_{n}
\tau^{p}_{\alpha\beta}-\tau^{p}_{\alpha\beta} a^{\beta}_{n}
=\eta^{p}_{\alpha\beta},
$$
where $a_{n}^{\alpha}\equiv\p\vec f^{\alpha}/\p\vec u_n^{\alpha}$, 
and $\eta^{p}_{\alpha\beta}$ is a $q_{\alpha}\times q_{\beta}$ matrix
whose entries are  differential polynomials in the entries of
the matrices \smash{$\tau^{j}_{\alpha\beta}$} with $j<p$, and in
the entries of coefficients of the formal series
$\mathfrak{F}^{\alpha}$, $\mathfrak{B}_{\alpha\gamma}$ and
$\mathfrak{T}_{\alpha\gamma}$ with $\gamma>\beta$.
\looseness=-1

Since the matrices $ a^{\alpha}_{n}$ have no common eigenvalues  by
assumption, we always can (see \cite{gant}) successively solve the
above equations with respect to
$\tau^{p}_{\alpha\beta}$ for $p=1,\!2,\dots$,
starting with the equations for ${\tau^{p}_{\alpha s}}$ and using
previously solved equations, if any occur.
What is more, the solution to these equations is unique
\cite{gant}. This completes the proof of the lemma.  $\square$

As an example, consider the system
\be\label{gkdv}
\ba{l}
u_t=(1-c)u_3+c v_3+3uu_1+3uv_1+3vu_1+3vv_1+g(w,w_1,w_2),\\
v_t=c u_3+(1-c)v_3+3uu_1+3uv_1+3vu_1+3vv_1+h(w,w_1,w_2),\\
w_t=w_3+w w_1,
\ea
\ee  
where $c$ is a constant.

The system 
$$
\ba{l}
u_t=(1-c)u_3+c v_3+3uu_1+3uv_1+3vu_1+3vv_1,\\
v_t=c u_3+(1-c)v_3+3uu_1+3uv_1+3vu_1+3vv_1,
\ea
$$
discovered by Foursov \cite{fo}, possesses a degenerate formal
symmetry of infinite rank\looseness=-1
$$
\mathfrak{L}_1=\left(
\ba{rr} 1 &-1\\
-1 & 1
\ea
\right)D^2,
$$
and for $c\neq \frac 12$ can be written in bi-Hamiltonian form in
infinitely many ways,
see \cite{fo}.
\looseness=-1

The equation $w_t=w_3+w w_1$ is nothing but the fabulous KdV equation, 
which has a nondegenerate formal symmetry of infinite rank 
$$
\mathfrak{L}_2=D^2+\frac 23 u + \frac 13 u_1 D^{-1},
$$
being in fact the recursion operator for this equation, see e.g.\
\cite{o}.
\looseness=-1

Thus, the requirements of Proposition~\ref{prop1} are met, and
(\ref{gkdv}) with arbitrary (smooth) functions $g$ and $h$ has a
degenerate formal symmetry of infinite rank and nonzero degree
$\mathfrak{R}=\mathfrak{T}^{-1}\mathfrak{L}\mathfrak{T}$,
where 
%
$\mathfrak{L}=\diag(\mathfrak{L}_{1},\!\mathfrak{L}_{2})$.

It would be interesting  to find out under which conditions
the system (\ref{gkdv}) has a {\em nondegenerate} formal symmetry of
infinite rank and nonzero degree, and we intend to analyse this
problem in more detail elsewhere.   

A fairly straightforward but quite useful application of 
Proposition~\ref{prop1} is given by the following result.
\looseness=-1
\begin{cor}\label{cor3}
Let $\vec f^{\alpha}=\sum
_{i=0}^{n} a_{i}^{\alpha}(x)\vec
u^{\alpha}_{i}$, $\alpha=1,\dots,s-1$, where $a_{i}^{\alpha}(x)$ are
$q_{\alpha}\times q_{\alpha}$ matrices. 
Denote for convenience $a_{n}^{s}=\p\vec{f}^s/\p\vec{u}^s_n$, and
suppose that the matrices $a_{n}^{\alpha}$ and 
$a_{n}^{\beta}$ have no common eigenvalues
(i.e., the eigenvalues in question
are distinct as functions)  
for all $\alpha\neq
\beta$, $\alpha,\beta=1,\dots,s$. 
Further assume that the evolution system $\vec u_t^s=\vec{f}^s$
possesses a formal symmetry
$\mathfrak{L}_{s}$ of infinite rank and nonzero degree, 
and at least one of the matrices $ a_{n}^{\alpha}$
is nonzero.\looseness=-1

Then the system {\rm (\ref{eveq3})} with arbitrary (smooth) functions
$\vec h^{\alpha}(x,t,\ \vec u^{\alpha+1}, \vec
u^{\alpha+1}_{1},\dots, 
\vec u^{\alpha+1}_{n-1},\dots,\vec u^s, 
\vec u^s_{1},\dots,\vec u^s_{n-1})$,
$\alpha=1,\dots,s-1$, also possesses a formal symmetry 
of infinite rank and nonzero degree.
\end{cor}

{\bf Proof.} We just take $\mathfrak{L}_{\alpha} =
\mathfrak{F}_{\alpha} = \sum_{i=0}^{n}a^{\alpha}_{i}(x)D^i$ for
$\alpha=1,\dots,s-1$. $\square$

\section{Existence of nondegenerate formal symmetries}

While applying  the existence of formal symmetry of infinite rank as 
an integrability test one usually requires  that the
system in question should have a {\em nondegenerate} formal
symmetry, cf.\ \cite{mik1}. The results that follow provide easily
verifiable sufficient conditions for the existence of formal symmetry
with this property.

\begin{cor}\label{cor4}
Under the assumptions of Corollary~\ref{cor3}, suppose that the system 
$\vec u_t^s=\vec{f}^s$ has a {\em nondegenerate} formal symmetry
$\mathfrak{L}_{s}$ of infinite rank and of nonzero degree $k$.
Further assume that at least one of the following conditions holds:
\looseness=-1
\begin{itemize}
\item[(i)] $\det  a_{\alpha}^n\neq 0$, $\alpha=1,\dots,s-1$; 
\item[(ii)] all matrices $ a_{\alpha}^i$, $\alpha=1,\dots,s-1$,
    $i=0,\dots,n$, are constant matrices;
\item[(iii)] one of the matrices $ a_{\alpha}^n$ 
    (say, $a_{\delta}^{n}$) is degenerate:
$\det a_{\delta}^{n}=0$; 
    $\det a_{\alpha}^n\neq 0$, $\alpha\neq\delta$,
    $\alpha=1,\dots,s-1$, and either  
 a) there exists $m\in\mathbb{N}$  such that $m<n$ and we have 
     $a_{\delta}^{m+1}=0,\dots,  a_{\delta}^{n}=0$ while 
     $a_{\delta}^{m}\neq 0$ and $\det a_{\delta}^{m}\neq 0$, or 
  b) all matrices  $ a_{\delta}^{j}$, $j=0,\dots,n$, are constant
ones.
\end{itemize}
\vspace{-1.5mm}

Then the system {\rm (\ref{eveq3})} with arbitrary (smooth) functions
$\vec h^{\alpha}
(x,t,\ \vec u^{\alpha+1}, \vec u^{\alpha+1}_{1},\dots,  
\vec u^{\alpha+1}_{n-1},\dots,\allowbreak \vec u^s, \allowbreak 
\vec u^s_{1},\dots, 
\vec u^s_{n-1})$, $\alpha=1,\dots,s-1$, possesses a 
{\em nondegenerate} formal symmetry $\mathfrak{R}$ of infinite rank
and nonzero degree.
\end{cor}
\vspace{-2mm}
\hspace*{\parindent}{\bf Proof.} In all cases we can represent
$\mathfrak{R}$ in the form 
$\mathfrak{R}=\mathfrak{T}^{-1}\mathfrak{L}\mathfrak{T}$, where 
$\mathfrak{L}$ solves (\ref{trans}), and the nondegeneracy
of 
$\mathfrak{L}$ clearly implies the same  property  for $\mathfrak{R}$.
Therefore, it suffices to construct a nondegenerate solution
$\mathfrak{L}$  of nonzero degree $r$ for (\ref{trans}). 
We shall exhibit such solutions for all cases (i)--(iii). Their
nondegeneracy will be obvious from the construction.

In the case (i) let $r$ be the least common multiple of $n$ and $k$,
$\tilde n=r/n$, $\tilde k=r/k$, and we set
$\mathfrak{L}=\diag(\mathfrak{F}_{1}^{\tilde n},\dots,
\mathfrak{F}_{s-1}^{\tilde n}, \mathfrak{L}_{s}^{\tilde k})$. 

Likewise, in the case (ii) we set
$\mathfrak{L}=\diag({\mathbf{1}}_{q_{1}} D^k,\dots,
{\mathbf{1}}_{q_{s-1}} D^k, \mathfrak{L}_{s})$, where
${\mathbf{1}}_{q_{\alpha}}$  is $q_{\alpha}\times q_{\alpha}$ unit
matrix.

In the case (iii, a) let  $r$ be the least common multiple of $n$, $m$
and $k$,  and we set 
$\mathfrak{L}=\diag(\mathfrak{F}_{1}^{\tilde
n},\dots,
\mathfrak{F}_{\delta-1}^{\tilde n}, 
\mathfrak{F}_{\delta}^{\tilde m},\allowbreak
\mathfrak{F}_{\delta+1}^{\tilde n},\dots, 
\mathfrak{F}_{s-1}^{\tilde n}, \mathfrak{L}_{s}^{\tilde k})$, where 
$\tilde n=r/n$, $\tilde m=r/m$,
$\tilde k=r/k$. 

Finally, in the case (iii, b), taking for $r$ the least 
common multiple of $n$ and $k$, 
we set $\mathfrak{L}=\diag(\mathfrak{F}_{1}^{\tilde n},\dots,
\mathfrak{F}_{\delta-1}^{\tilde n},\allowbreak 
{\mathbf{1}}_{q_{\delta}}D^r,
\mathfrak{F}_{\delta+1}^{\tilde n},\dots,
\mathfrak{F}_{s-1}^{\tilde n}, \mathfrak{L}_{s}^{\tilde k})$,
where ${\mathbf{1}}_{q_{\delta}}$  is $q_{\delta}\times q_{\delta}$ 
unit matrix, 
$\tilde n=r/n$, $\tilde k=r/k$. 
$\square$

\begin{cor}\label{cor5}
Let 
$
\vec f^{\alpha}=\sum_{i=0}^{n} a_{i}^{\alpha}(t)\vec u^{\alpha}_{i}
$, 
$\alpha=1,\dots,s-1$, where $a_{i}^{\alpha}(t)$ are
\hbox{$q_{\alpha}\times q_{\alpha}$} matrices. 
Again denote for convenience $a_{n}^{s}=\p\vec{f}^s/\p\vec{u}^s_n$,
and suppose that the matrices $a_{n}^{\alpha}$ and 
$a_{n}^{\beta}$ have no common eigenvalues, 
i.e., the eigenvalues in question
are distinct as functions, 
for all $\alpha\neq
\beta$, $\alpha,\beta=1,\dots,s$, 
and at least one of the matrices $ a_{n}^{\alpha}$
is nonzero.

Then the system {\rm (\ref{eveq3})} with arbitrary smooth functions
$\vec h^{\alpha}(x,t,\ \vec u^{\alpha+1},\ \vec u^{\alpha+1}_{1},
\dots,\ \vec u^{\alpha+1}_{n-1},\dots,\ \vec u^s,\allowbreak\vec
u^s_{1},
\dots,\ \vec u^s_{n-1})$, $\alpha=1,\dots,s-1$, possesses a
(nondegenerate) formal symmetry $\mathfrak{L}_{s}$ of infinite rank
and of nonzero degree $k$, if so does the system $\vec
u_t^s=\vec{f}^s$.
\looseness=-1
\end{cor}

{\bf Proof.} 
Let $\mathfrak{L}=\diag({\mathbf{1}}_{q_{1}} D^k,\dots,
{\mathbf{1}}_{q_{s-1}} D^k, \mathfrak{L}_{s})$ and $\mathfrak{T}$ be 
a formal series constructed in Lemma~\ref{lemma}. Then 
$\mathfrak{R}=\mathfrak{T}^{-1}\mathfrak{L}\mathfrak{T}$ is a 
formal symmetry of infinite rank and of degree $k\neq 0$ for
(\ref{eveq3}). Finally, if $\mathfrak{L}_{s}$ is nondegenerate,
then so does $\mathfrak{R}$. $\square$
\looseness=-1

For instance, the Bakirov system (\ref{bakir}) and the system
(\ref{svk}), investigated by Sanders and van der Kamp \cite{sv},
indeed meet the requirements of Corollaries~\ref{cor3}, \ref{cor4}
and~\ref{cor5}  and therefore  have nondegenerate formal symmetries
of infinite rank and nonzero degree. What is more, by
Corollary~\ref{cor5} any system of the form
$$
u_{t}=a(t) u_{n}+K(x,t,v,v_1,\dots,v_{n-1}),\\
v_{t}=b(t) v_n 
$$
has a nondegenerate
formal symmetry of infinite rank and nonzero degree, provided 
$a(t)\neq b(t)$.\looseness=-1

Following Kupershmidt \cite{bk}, consider a system $\vec u_t=\vec
F(x,t,\vec u,\dots,\vec u_n)$ with $n\geq 2$ and $\det\p\vec{F}/\p
\vec u_n\neq 0$, and its (vectorial) {\em logarithmic extension} 
\be\label{log}
\vec u_{t}=\vec F(x,t,\vec u,\dots,\vec u_n),\\
\vec v_{t}=\vec G(x,t,\vec u,\dots,\vec u_{n-1}).
\ee
Here $\vec u$ and $\vec v$ are $q_1$- and $q_2$-component vectors, 
respectively, and $\vec G$ is an arbitrary (smooth) $q_2$-component
vector function.

By Corollary~\ref{cor5} the system (\ref{log}) possesses a
(nondegenerate) formal  symmetry of infinite rank and nonzero
degree if so does $\vec u_t=\vec F(x,t,\vec u,\dots, \vec u_n)$. 
This fact suggests that, in
addition to the four types of extensions of integrable systems,
introduced in
\cite{bk},  it is natural to consider the fifth one, namely, the
extensions  which ``inherit'' (nondegenerate) formal symmetry from
the original system.

\section{Conclusions and discussion}

We have shown above that a fairly large class
of evolution systems~(\ref{eveq3}) has a
(non\-de\-ge\-nerate)  formal symmetry of infinite rank and nonzero
degree, provided so do all ``building blocks'' of (\ref{eveq3}),
that is, $\vec u^\alpha_t=\vec f^\alpha$, and the
coefficients of formal symmetries  of the first $s-1$ blocks depend
on $x$ and $t$ only. In other words, 
under certain conditions the system (\ref{eveq3}) inherits some of
integrability properties of its blocks.
\looseness=-1


Let us also mention that once a solution $\vec u^s(x,t)$ of 
$\vec u^s_t=\vec f^s$ 
is known,  recovering the corresponding solution of (\ref{eveq3}) 
amounts to solving linear inhomogeneous PDEs, provided 
$\vec f^{\alpha}$,
$\alpha=1,\dots,s-1$ are linear in $\vec u^{\alpha}_j$. In this case,
if the system $\vec u^s_t=\vec f^s$ is exactly solvable, then the 
same is true for (\ref{eveq3}). However, as show the
examples of the Bakirov system \cite{bak91}, and of the systems
constructed by Sanders and van der Kamp in \cite{sv}, if the system
$\vec u^s_t=\vec f^s$ has infinitely many non-Lie-point  
local generalized symmetries, the system (\ref{eveq3}) does not
necessarily have the same property even if it possesses a formal
symmetry of infinite  rank and nonzero degree. We encounter here
an intriguing phenomenon of 
`disappearing' of symmetries, which, surprisingly, is due to some
subtle number-theoretical effects \cite{bsw,sv}.
\looseness=-1

\subsection*{Acknowledgements}
It is my great pleasure to express deep gratitude to organizers of
$8^{\mathrm{th}}$ ICDGA for the opportunity to present my results at
this conference and  for their kind hospitality. I also thank very
much  Prof. B.A. Kupershmidt for kindly sending 
me the preprint version of \cite{bk}. Last but not least, I am
sincerely grateful  to Dr.~E.V.~Ferapontov and Dr.~M.~Marvan 
for the fruitful discussions on the results of the present paper.
\looseness=-1

This research was supported by DFG via Graduiertenkolleg 
``Geometrie und Nichtlineare Analysis'' of Institut f\"ur Mathematik of
Humboldt-Universit\"at zu Ber\-lin, Germany, where the author held  a
postdoctoral fellowship. I also acknowledge the partial support from the
Ministry of Education, Youth and Sports of Czech Republic under Grant
MSM:J10/98:192400002,  and from the Czech Grant Agency under Grant 
No.~201/00/0724.
\looseness=-1


\end{document}